\documentclass[aps,pra,twocolumn,showpacs,groupedaddress]{revtex4}
\usepackage{graphicx}
\usepackage{amsmath}

\begin{document}

\newcommand{\Singlet}{X^1\Sigma^+}
\newcommand{\Triplet}{a^3\Sigma^+}

\title{Isotopic difference in the heteronuclear loss rate in a
two-species surface trap}
\author{M. E. Holmes, M. Tscherneck, P. A. Quinto-Su, and  N. P. Bigelow}
\address{Department of Physics and Astronomy, The Laboratory for Laser Energetics,
and The Institute of Optics\\ The University of Rochester,
Rochester, NY 14627}

\date{\today}

\begin{abstract}
We have realized a two-species mirror-magneto-optical trap
containing a mixture of $^{87}$Rb ($^{85}$Rb) and $^{133}$Cs
atoms.  Using this trap, we have measured the heteronuclear
collisional loss rate $\beta_{Rb-Cs}'$ due to intra-species cold
collisions.  We find a distinct difference in the magnitude and
intensity dependence of $\beta_{Rb-Cs}'$ for the two isotopes
$^{87}$Rb and $^{85}$Rb which we attribute to the different
ground-state hyperfine splitting energies of the two isotopes.

\end{abstract}

 \maketitle

Since the first experiments in 1999~\cite{atom_micro}
demonstrating the trapping of atoms using micron scale wires
fabricated on a reflective substrate (the so-called ``atom chip")
there has been increased interest in techniques for cooling,
trapping, and controlling atoms at surfaces.  Various devices have
been proposed and are now being realized in the laboratory
including atomic beamsplitters and wave
guides~\cite{beam_splt_guide,guide_neut,two_wguide,appl_integ_micro}.
Indeed, sufficient progress has been made that a BEC can now be
created and manipulated using atom-chip techniques.  In a parallel
development, the investigation of cold atomic clouds of mixed
atomic species has also attracted substantial attention, giving
rise to intense efforts to generate cold hetero-nuclear
molecules~\cite{rbcs_mol,quan_computer_mol} for application to
fundamental measurements \cite{two_wguide} and for quantum
information technologies \cite{quan_computer_mol}.  To date,
however, there have been no reports of mixed species trapping with
surface trap technologies.

In this paper we report the realization of a two-species surface
trap -- the two-species mirror-magneto-optical trap (TSMMOT).  As
with other atom traps, the performance of the trap (e.g density,
number, etc.) is strongly affected by collisionally induced trap
loss ~\cite{two_spec_nacs,hetero_hsc_coll,tl_two_spec_rbcs}.  We
have therefore used the TSMMOT to investigate the cold collisions
of atomic Cs with $^{87}$Rb ($^{85}$Rb).  We focus on measurements
made in the low intensity regime ($2 I_{sat} \le I \le 8 I_{sat}$
where $I_{sat}$ is the atomic trapping transition saturation
intensity). Mixed species Cs-Rb trap losses have been recently
characterized in a standard MOT over a broad range of laser
intensities~\cite{tl_two_spec_rbcs}. Our results are distinct in
that we find an isotopic difference, which to our knowledge has
not been previously observed, and which we attribute to
ground-state interspecies hyperfine changing processes.

In our experiments, trapping light was provided by line-narrowed
extended cavity diode lasers locked to the trapping transitions
using a dichroic (DAVLL) scheme
~\cite{laser_frq_zeeman,freq_stble_diode}. Acousto-optical
modulators (AOM) were used to detune the light from the locking
point to the cooling transition by -1 $\Gamma_{Rb}$ (-2.1
$\Gamma_{Rb}$) for
 $^{85}$Rb ($^{87}$Rb) and by -1.3 $\Gamma_{Cs}$ for Cs.
To assure uniform Gaussian beams, all trapping light was passed
through single-mode optical fibers. After the fibers, the beams
had an $1/e^2$ waist of 0.4 cm. A series of polarizing beam
splitter cubes (PBSC) and half-wave plates were used to mix the
trapping light and to tune the individual intensities of the
repumping and trapping light of each species. The adjustments were
made to optimize the size, shape, and overlap of both atomic
clouds.

The surface used for this TSMMOT configuration was fabricated
in-house using thin-film hybrid technology~\cite{atom_micro}: a
top layer 0.7 $\mu$m thick of highly reflective (95\%) Ag was
evaporated onto a 1 $\mu$m SiO$_{2}$ sputtered layer, all
deposited onto a 300 $\mu$m thick Si wafer.  Similar techniques
were used to pattern chips capable of magnetic surface trapping.
The typical TSMMOT hovered \mbox{$\sim$3.5 mm} above the
mirror-surface to ensure that surface effects play no role. Taking
the geometry into account~\cite{eso_calc}, the maximum total
intensity within the TSMMOT region is 13 mW/cm$^{2}$ for Rb and 53
mW/cm$^{2}$ for Cs. A set of anti-Helmholtz coils typically
produced a magnetic field gradient of about 40 G/cm. Three
orthogonal Helmholtz pairs (``trim coils"), were used to
compensate for residual stray magnetic fields and to guarantee
good overlap between the two atomic species. The pressure in the
chamber was maintained at $\le10^{-8}$ Torr. The atoms were
introduced using getter sources placed inside the vacuum chamber
approximately 5 cm from the B-field minimum~\cite{load_getter}.

\begin{table*}
\caption{\label{tab:parameter} Experimental parameters used to
characterize $^{85}$Rb and $^{87}$Rb with Cs.}
\begin{ruledtabular}
\begin{tabular}{cccc}

Parameter& $^{85}$Rb &  $^{133}$Cs& $^{87}$Rb\\ \hline

Cooling transition & $5S_{1/2} (F=3)$ & $6S_{1/2} (F=4)$ &
$5S_{1/2} (F=2)$\\

&$\rightarrow 5P_{3/2} (F'=4)$ & $\rightarrow 6P_{3/2} (F'=5)$&
$\rightarrow 5P_{3/2} (F'=3)$\\

Natural linewidth $\Gamma$ [MHz] &2$\pi$ x 5.98 &2$\pi$ x
5.22&2$\pi$ x 6.07\\

Saturation intensity $I_{sat}$ [mW/cm$^2$] & 1.64& 1.10&1.67\\

Detuning from cooling transition &-1.0 $\Gamma_{Rb}$&-1.3
$\Gamma_{Cs}$&-2.1 $\Gamma_{Rb}$\\

Total intensity I$_{tot}$ [mW/cm$^2$] &4 - 13&53&4 - 13\\

Atom number $N$& $1 \times 10^{7}$ - 2$\times 10^{7}$&$1 \times
10^{7}$& 3$\times 10^{6}$ - 9$\times 10^{6}$\\

vertical waist with (without) Cs [$\mu$m] &215 - 300 (260 -
330)&447&250 - 395 (340 - 845)\\

horizontal waist with (without) Cs [$\mu$m] &420 - 470 (420 -
570)&169&115 - 150 (130 - 170)\\

peak density $n$ with (without) Cs [cm$^{-3}$]&2$\times 10^{11}$
(3 - 4.5$\times 10^{11}$)&1.5$\times 10^{11}$&6$\times 10^{10}$
(4$\times 10^{10} - 1\times 10^{11}$)\\

\end{tabular}
\end{ruledtabular}
\end{table*}

\begin{figure}
\includegraphics[width=9.0cm]{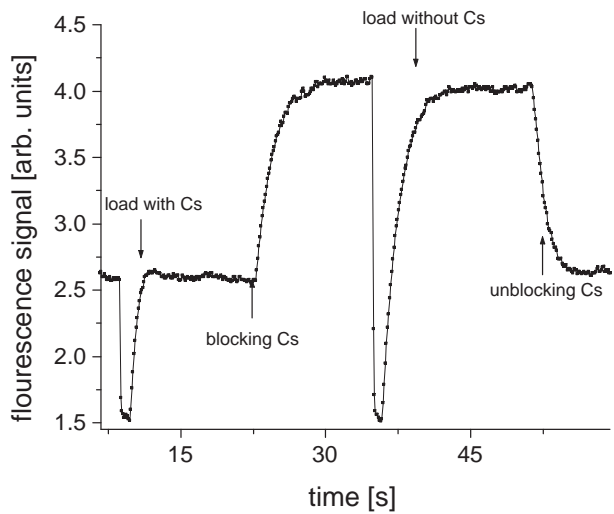}
\caption{\label{fig:percloss} The transient loading signal
of $^{87}$Rb , both with and without Cs, as a function
of time.}
\end{figure}

To assure full three-dimensional overlap of the two trapped
species, the clouds were imaged with a pair of CCD cameras aligned
on separate axes perpendicular to the chip. A third
high-performance CCD (high linearity) was used to image the MOTs
and measure their spatial distributions. The shape of the two
clouds was that of an oblate spheroid. The measured waists are
noted in Table~\ref{tab:parameter}.  The number of trapped atoms
was determined by measuring the fluorescence using two calibrated
photo-detectors combined with narrow-band interference filters
(bandwidth of about 9 nm) capable of isolating the fluorescence of
the individual atomic species. The total number of trapped atoms
was typically 2$\times 10^{7}$ (9$\times 10^{6}$) for $^{87}$Rb
($^{85}$Rb) and $1\times 10^{7}$ for Cs. This along with our waist
measurements yields peak densities of 4.5$\times 10^{11}$
($10^{11}$) atoms/cm$^{3}$ for Rb and 1.5$\times 10^{11}$
atoms/cm$^{3}$ for Cs.

The amount of collision induced trap-loss depended critically on the overlap between the
two atomic clouds. During all experiments, the overlap was $\gtrsim95\%$ by volume.

The Cs MOT was imposed onto the Rb MOT by controlled blocking and
unblocking of the repump light for Cs. Fig.~\ref{fig:percloss}
shows the loading behavior of the Rb atoms in the absence and
presence of Cs. Losses
 as large as 78\% in the Rb atom number due to Cs were observed.  By fitting
this data to a transient loading rate
equation~\cite{two_spec_nacs,tl_two_spec_rbcs}, the heteronuclear
loss rate was obtained. The dependence of trap loss on the total
intensity of the Rb trap laser was measured by keeping the Cs
laser intensity and the number of Cs atoms in the trap constant.

The collisions of atoms in a MOT can be characterized by the
interplay of the loading and loss rate of trapped atoms.  The
time-dependent rate equations that model this process can be
written:
\begin{align}
 \label{eq:compload}
 \begin{split}
d_{t}N_{Rb}(t)=\tau_{Rb}&-\gamma_{Rb}N_{Rb}(t)-\beta_{Rb-Cs}'n_{Cs}N_{Rb}(t)\\&-\beta_{Rb-Rb}n_{Rb}N_{Rb}(t)
\end{split}
\end{align}
where $N_{Rb}$ is the number of Rb atoms, $\tau_{Rb}$ is the trap
filling rate, $\gamma_{Rb}$ is the loss coefficient due to
background collisions, $\beta_{Rb-Rb}$ is the loss rate due to
homonuclear collisions of atoms of one species, and
$\beta_{Rb-Cs}'$ is the loss rate due to heteronuclear collisions
of atoms between the two species. The atom number densities,
$n_{Cs}$ and $n_{Rb}$, are experimentally defined as the peak
number of atoms divided by the total volume, calculated using the
Gaussian $1/e^2$ waists. Similar to previous treatments in the
literature, we assume that we are in the density-limited
regime~\cite{tl_two_spec_rbcs}.

In our experiments, we observe that the trapped atom number and
density for the Cs trap is essentially undisturbed by the
introduction of Rb atoms into the trap.  Hence in our analysis we
treat the Cs density as a constant.  By contrast, the number of
trapped Rb atoms is dramatically affected by the presence of Cs
atoms in the trap.

We begin our analysis by assuming that $\beta_{Rb-Cs}' \gg
\beta_{Rb-Rb}$~\cite{tl_two_spec_rbcs,iso_diff_rb}. Eq.
\ref{eq:compload} can then be written as
\begin{align}
 \label{eq:dispersion}
 \begin{split}
d_{t}N_{Rb}(t) &=
\tau_{Rb}-(\gamma_{Rb}+\beta_{Rb-Cs}'n_{Cs})N_{Rb}(t)
\\
&= \tau_{Rb}-\gamma_{Rb}'N_{Rb}(t)
\end{split}
\end{align}
where $\gamma_{Rb}'$  is the total loss rate of the mixed trap.
For the parameters of our experiment (chamber pressure etc.), even
in the absence of Cs, $\beta_{Rb-Rb}$ can be neglected and Eq.
\ref{eq:compload} becomes
\begin{equation}
\label{eq:pureRb} d_{t}N_{Rb}(t)=\tau_{Rb}-\gamma_{Rb}N_{Rb}(t).
\end{equation}
Combining Eqs. \ref{eq:dispersion} and \ref{eq:pureRb},
$\beta_{Rb-Cs}'$ is given by
\begin{equation}
\label{betaprime}
\beta_{Rb-Cs}'=\frac{(\gamma_{Rb}'-\gamma_{Rb})}{n_{Cs}}.
\end{equation}

\begin{figure}
\includegraphics[width=8cm]{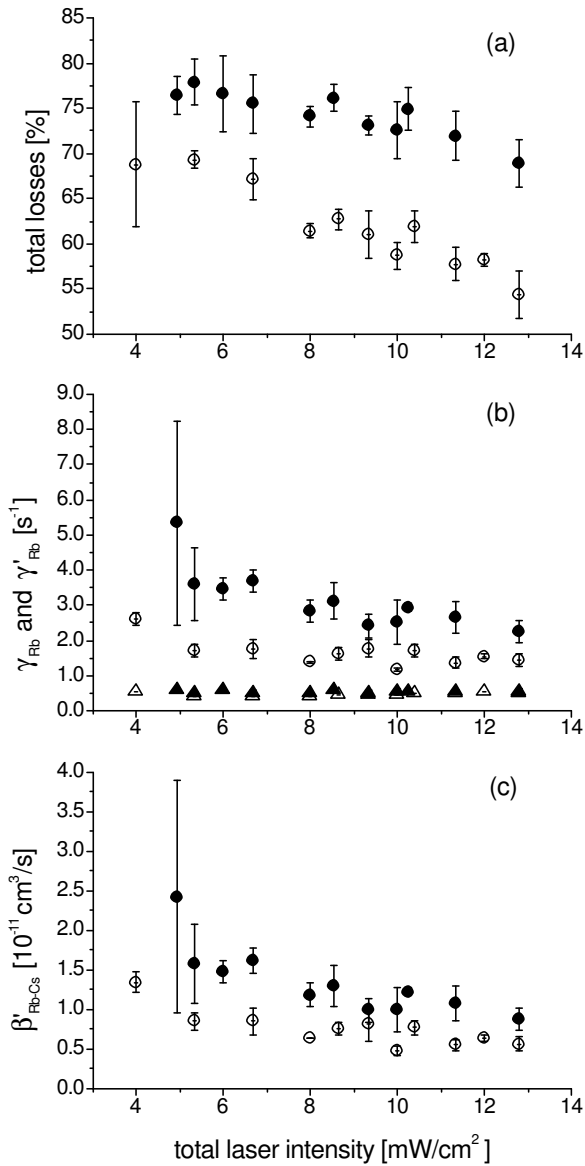}
\caption{\label{betaloss}Trap losses for both isotopes as a
function of Rb laser intensity. The solid and hollow symbols in
all plots represent $^{87}$Rb and $^{85}$Rb, respectively. Plot
(a) shows the overall losses in percent, plot (b) the total loss
rates $\gamma_{Rb}'$ (circles) and $\gamma_{Rb}$ (triangles), with
and without Cs, and plot (c) $\beta_{Rb-Cs}'$.}
\end{figure}

Fig.~\ref{betaloss}a shows the measured average losses of Rb due
to Cs. The error bars correspond to standard deviations in mean
value, averaged over repeated experiments performed while keeping
experimental parameters constant. The losses decrease almost
linearly with increasing Rb laser intensity. They also show a
distinct isotopic difference: the losses for $^{87}$Rb are greater
than those for $^{85}$Rb. This behavior is also seen in the total
loss rate which is shown in Fig.~\ref{betaloss}b where we plot the
total loss rate $\gamma_{Rb}$ (for pure Rb) and the total loss
rate $\gamma_{Rb}'$ (for Rb+Cs) on the same graph. We observe no
change in $\gamma_{Rb}$ (pure Rb trap) for this intensity regime
and see no isotopic difference. This is consistent with previous
work~\cite{iso_diff_rb,trp_loss_param_rb}. However, the total loss
rate $\gamma_{Rb}'$ changed dramatically in the presence of Cs.
Again, the averaged loss rate for $^{87}$Rb is consistently higher
than for $^{85}$Rb. This isotopic difference is transferred onto
$\beta'_{Rb-Cs}$ (Fig.~\ref{betaloss}c) which was calculated using
Eq.~\ref{betaprime}.

We note that the absolute values of $\beta_{Rb-Cs}'$ have
significant uncertainties ($\sim 50\%$) which arise from
systematics in determining the exact atom number (not included in
our error bars). However, this uncertainty is identical for each
isotope of Rb and hence does not change the observed isotopic
difference.

We attribute the isotopic difference to the difference in the
hyperfine ground-state splitting energy of the two isotopes of
Rb~\cite{iso_diff_rb,trp_loss_param_rb}.  Experimentally, isotopic
difference in trap loss due to ground state hyperfine structure
were first observed in experiments performed using pure Rb
traps~\cite{iso_diff_rb}.  Phenomenologically, the effect was
explained by noting that the trap depth, which decreases with
decreasing intensity, is approximately the same for the two
isotopes whereas the ground-state hyperfine splitting energy is
smaller for $^{85}$Rb than for $^{87}$Rb. As hyperfine changing
collisions involving $^{87}$Rb release more energy than those
involving $^{85}$Rb they cause more trap loss in the low intensity
regime. In particular, as a function of the trap laser intensity,
$\beta_{Rb-Rb}$ has been observed to decrease with increasing
intensity, reach a minimum and increase again for higher
intensities.  For an ideally aligned MOT, the minimum is reached
when the trap depth equals the hyperfine splitting energy. As the
hyperfine energies for the two Rb isotopes are different, this
minimum occurs at different trap intensities. For a slightly
misaligned MOT this minimum is shifted to higher intensities, but
the shape of the curves and the isotopic difference is preserved.

We find that the behavior of $\beta'_{Rb-Cs}$ parallels the
homonuclear Rb experiments. In the low intensity regime, we see a
decrease of $\beta'_{Rb-Cs}$ with increasing intensity
\cite{two_spec_nacs,tl_two_spec_rbcs}. In addition, the slope of
the curve is found to be smaller for $^{85}$Rb than for $^{87}$Rb,
as the hyperfine splitting energy of $^{85}$Rb is smaller and
therefore the minimum is reached at lower intensities.

 We note that ground-state heteronuclear hyperfine
changing collisions have also been observed in mixtures of sodium
and rubidium in our labs~\cite{hetero_hsc_coll} however that work
was not performed in the environment of a surface trap.

In summary we presented heteronuclear trap loss measurements in a
mixed Rb-Cs TSMMOT. At low intensites, there is an isotopic
difference between $^{85}$Rb and $^{87}$Rb. Our loss measurements
agree well with previous data obtained for a mixed Rb-Cs trap,
however no isotopic differences were reported in that work. With
well overlapped cloud centers, losses up to ~78\% can be obtained.
To our knowledge this is the highest loss reported for a mixed
Rb-Cs MOT.

\begin{acknowledgments}
We would like to thank Michael Wulf and Mark J. Feldman for
providing the mirror surface. We are grateful for contributions
from Jim Steinman, Laura Pickel, and Jeremy Weeden. This work was
supported by the National Science Foundation, the Office of Naval
Research and the Army Research Office.
\end{acknowledgments}

\bibliography{RbCsNew}

\begin{thebibliography}{16}
\bibitem[{eso()}]{eso_calc}
\bibinfo{note}{In our calculation of the excited state occupation, the total
  beam power was computed with a total of six beams within the MMOT region.
  Also, the trap offset from the center of the $1/e^2$ beam waist was carefully
  included.}
\end{thebibliography}

\end{document}